\documentclass[pre,twocolumn,unsortedaddress,floatfix,amssymb]{revtex4}
\usepackage{times}
\usepackage{graphicx}
\usepackage{amsmath}
\usepackage{color}
\usepackage{natbib}
\bibpunct{[}{]}{,}{n}{}{;}

\DeclareGraphicsExtensions{.eps, .eps.bz2}

\newcommand{\br}{\textbf{r}}

\begin{document}

\title{Pore-polymer interaction reveals non-universality in forced polymer 
translocation}

\author{V. V. Lehtola}
\author{K. Kaski}
\author{R. P. Linna}
\affiliation{
  Department of Biomedical Engineering and Computational Science,
  Aalto University, P.O. Box 12200, FI-00076 Aalto, Finland
  }

\date{August 17, 2010}


\begin{abstract}
We present a numerical study of forced polymer translocation by using two 
separate pore models. Both of them have been extensively used in previous 
forced translocation studies. We show that variations in the pore model affect 
the forced translocation characteristics significantly in the biologically 
relevant pore force, i.e. driving force, range. Details of the model are shown 
to change even the
obtained scaling relations, which is a strong indication of strongly 
out-of-equilibrium dynamics in the computational studies which have 
not yet succeeded in addressing the characteristics of the forced translocation 
for biopolymers at realistic length scale.
\end{abstract}

\maketitle

\section{Introduction}

Polymer translocation has been under intensive research for the past
decade due to its relevance for {\it e.g.} ultra-fast DNA
sequencing~\cite{kasianowicz, storm,li,meller_prl,zwolak} and many 
biological processes~\cite{alberts}. Most of the computational
research on forced translocation, where the polymer is driven through
a nanoscale pore by a potential, has dealt with fairly weak pore
potential or force and assumed close-to-equilibrium dynamics.
Recently, hydrodynamics has been shown to significantly affect forced
translocation for experimentally and biologically relevant force
magnitudes~\cite{fyta_pre,ourepl}. The pore geometry has been noted to
have effect on the experimental translocation process (see
e.g.~\cite{zwolak}). In addition, the effect of the pore model has
been discussed in~\cite{lubensky, slonkina}. However, the significance
of the pore model in the forced polymer translocation has not yet been
determined.

We have recently made a comparison of the unforced and forced 
translocation~\cite{ourunf}. In the first case, the process was seen to be 
close to equilibrium, as expected, and accordingly its dynamics turned out to 
be robust against variations 
in the computational model. However, in the second 
case the observed out-of-equilibrium characteristics for biologically and 
experimentally relevant pore force magnitudes, made the dynamics more sensitive 
to differences in the computational models~\cite{ourepl, ourpre}. Accordingly, 
we expect the pore model to play a significant role 
in the forced translocation case~\cite{ourunf}.

So far, we have consistently used a cylindrical pore implemented 
by a potential symmetrical around the pore axis~\cite{ourepl,ourpre,ourunf}. 
However, this  
differs from the typical pore implementation where the pore is formed by 
removing one particle from a wall consisting 
a monolayer of immobile, point-like 
particles~\cite{exponents,luo_slowfast,gauthier,fyta,fyta2}. 

A notable difference is that the latter pore implementation results 
in a driving potential that is inhomogeneous in the direction of the pore axis. 
In the case of lattice Boltzmann simulations these bead-pore models include 
both 
the square~\cite{fyta,fyta2}, and cylindrically~\cite{bernaschi} shaped pores. 
Typically, the pore-polymer interaction is solely repulsive but the effect of 
attractive interaction has been investigated in a 2D model~\cite{luo_prl}. 

In order to investigate how susceptible translocation dynamics is to 
variations in the pore model we have implemented in our computational 
translocation model also the bead pore used in~\cite{luo_slowfast}. We compare 
translocation processes for the bead pore and our original cylindrical pore. 
The model dependency is expected to be strongest in the pore region
under strong forcing. We also try to evaluate the effect of the pore model for 
moderate forcing. 

When the force bias inside the pore is small, the velocities involved are small 
enough 
for the hydrodynamics to be neglected~\cite{ourepl,ourunf}, and the Langevin 
dynamics (LD) is a 
valid choice for the computational model. Assuming that the 
polymer remains close to equilibrium throughout the translocation, the control 
parameter may be defined 
as the ratio of the total external force and the solvent
friction, $\zeta = f_{tot}/ \xi$ as done by Luo et al. in ~\cite{luo_slowfast}. Then the 
translocation dynamics would depend only on $\zeta$. They reported 
this to be the case even for a long polymer chain driven through the pore with 
a significant value of total pore force $f_{tot}$. The scaling of the 
average 
translocation time with respect to the chain length, $\tau \sim N^\beta$ would 
then be determined solely by $\zeta$. This 
is in clear contradiction with our earlier observation that the scaling 
exponent $\beta$ increases with the force 
due to crowding of the monomers on the {\it trans} side~\cite{ourepl,ourpre}. 
In addition, the effect of the pore model along with other model-dependent 
factors would in this case be negligible~\cite{ourunf}.

The increase of $\beta$ with force cannot hold asymptotically as pointed out 
in~\cite{luo_slowfast}, since ultimately very long polymers would translocate 
faster with a smaller pore force. From Fig.~1~b) in~\cite{ourepl} it can be 
evaluated that this unphysical condition would be seen for 
$N \gtrsim 10^6$ for the pore force magnitudes $1$ and $10$. This is comparable 
to actual DNA lengths of $10^6 \ldots 10^9$ that are far beyond polymer lengths 
of $N \lesssim 1000$ used in simulations. The translocation dynamics then would 
have to be different for $N > 10^6$, which is in keeping with our 
earlier qualitative description of the forced 
translocation~\cite{ourepl,ourpre}. The polymers were seen to be driven out of 
equilibrium throughout the translocation already for modest pore force 
magnitudes. The continually increasing drag force on the {\it cis} and crowding 
on the {\it trans} side were seen to change the force-balance condition, which 
is directly reflected on the translocation dynamics. This force balance is sure to 
be different for very long polymers.

The suggested control parameter $\zeta$, if applicable, would imply
that the forced translocation could be completely characterized
through varying it, which again is in stark contrast to our 
observation that the forced translocation is a non-universal, 
out-of-equilibrium process for biologically relevant magnitudes of
pore force~\cite{ourepl,ourpre}. Therefore, it is essential to 
determine if this contradiction arises from using different pore
models. 

This paper is organized such that we first describe the polymer model
in Section~\ref{pol_mod} and the translocation models in
Section~\ref{tra_mod}. The relation between computational and physical
parameters is discussed in Section~\ref{phys_par}.  The results are 
presented and discussed in Section~\ref{res}. We conclude the paper by
summarizing our findings in Section~\ref{sum}. 

\section{Polymer model}
\label{pol_mod}

The standard bead-spring chain is used as a coarse-grained polymer
model with the Langevin dynamics. In this model adjacent monomers
are connected with an-harmonic springs, described by the  finitely
extensible nonlinear elastic (FENE) potential,
\begin{equation}
U_{FENE} = - \frac{K}{2} R^2 \ln \big ( 1- \frac{r^2}{R^2} \big ),
\label{fene_mc}
\end{equation}
where $r$ is the length of an effective bond and $R = 1.5$ is the maximum bond
length. The Lennard-Jones (LJ) potential
\begin{eqnarray}
U_{LJ} &=& 4 \epsilon \left[ \left(\frac{\sigma}{r}\right)^{12} -
\left(\frac{\sigma}{r}\right)^{6} \right]
, \: r \leq 2^{1/6} \sigma \nonumber \\
U_{LJ} &=& 0 , \; r > 2^{1/6} \sigma,
\label{lj_md}
\end{eqnarray}
is used between all beads of distance $r$ apart. The parameter values
were chosen as $\epsilon = 1.2$, $\sigma = 1.0$ and $K = 60 / \sigma^2$. The 
used LJ potential mimics good solvent.

\begin{figure}
\centerline{
\includegraphics[angle=0,width=0.45\textwidth]{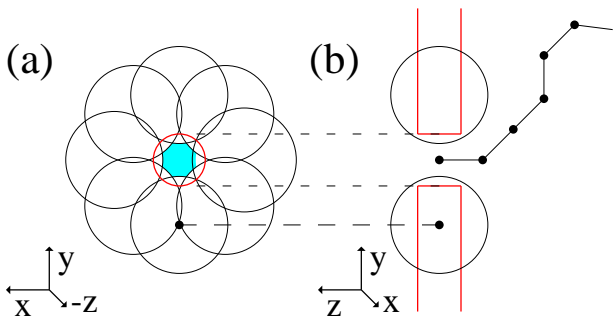}
}
\caption{
(Color online) Schematic depiction of the two pore models. (a) The pores are 
viewed from the {\it trans} side along the $z$-axis. The small red circle 
depicts the cylindrical pore of diameter $1.2 \sigma$. The bead pore is defined 
by the eight beads each at distance $1.5\sigma$ from the $z$-axis. The pore 
beads are drawn with circles using the LJ potential cutoff length $2^{1/6}$ as 
their radius. The light blue area in the center of the pore indicates the 
region where polymer beads have no interaction with the pore beads. In 
contrast, the cylinder pore model has a damped-spring-like potential that acts 
on particles everywhere inside the pore. 
(b) Side view. The polymer about to translocate ($s=1$) is drawn as connected 
dots. The potentials of the two pore models differ in both the $xy$-plane and 
along the $z$-axis.
} 
\label{pore_fig}
\end{figure}
\section{Translocation models}
\label{tra_mod}

To perform simulations of polymer translocation
in 3D we use our translocation model based on Langevin
dynamics~\cite{ourunf}. Hence, the time derivative of the momentum of bead $i$ 
reads as
\begin{equation}
 \dot{\mathbf{p}}_i(t) = - \xi \mathbf{p}_i(t) + \eta_i(t) + f(\br_i),
\label{langevin}
\end{equation}
where $\xi$, $\mathbf{p}_i(t)$, $\eta_i(t)$, and $f(\br_i)$ are the 
friction constant, the momentum, random force of the bead $i$, and the
external driving force, respectively. $f(\br_i)$ is constant
but exerted only inside the pore. For unforced translocation
$f(\br_i)=0$. $\xi$ and $\eta_i(t)$ are related by the
fluctuation-dissipation theorem. The dynamics was implemented by using
velocity Verlet algorithm~\cite{vangunsteren}. 

In our model the wall containing the pore was implemented as a 
surface on which no-slip boundary conditions are applied to the polymer beads.
The two different computational pore models to be compared can be  described as 
follows (see Fig.~\ref{pore_fig}):

{\it Cylindrical pore.} The pore, aligned with the $z$-axis, is of
diameter $1.2 \sigma (= 1.2 b)$ and length $l = n b$, where $b=1$ is the Kuhn 
length of the
model polymer, and $n$ is either $1$ or $3$. The pore is
implemented by a cylindrically symmetric damped harmonic potential
that pulls the beads toward the pore axis passing through the middle
of the pore in the $z$-direction. The pore is $n b$ long, so the total
pore force is taken as $f_{tot}= n f$, where $f$ is the
external driving force applied on each polymer bead inside the pore.

{\it Bead pore.} An octagonal composition of eight immobile particles was used 
as the bead pore model, as depicted in Fig.~\ref{pore_fig}. This pore model 
includes a 
region where a polymer bead can reside without interacting with the pore beads,
see Fig.~\ref{pore_fig}. The effective pore length is approximately $b$, so 
$f_{tot} =f$. Unlike in our 
model, in~\cite{luo_slowfast} also the wall was made of immobile particles. 
We verified that this has no effect on the translocation dynamics by 
reproducing essentially identical results 
on $\tau$ for the force magnitudes that were reported in~\cite{luo_slowfast}.

In our previous Langevin dynamics simulations with the cylindrical 
pore~\cite{ourepl,ourunf}, parameter values $\xi = 0.73$, $m = 16$,
and $k T = 1$ were used for the friction constant, the polymer bead
mass, and the temperature in reduced units, respectively. Hence, for long 
times the one-particle
self-diffusion constant was obtained from Einstein's relation as $D_0 =
k T/ \xi m \approx 0.086$. Time steps of $0.01$ and $0.03$ were
(previously) used in the forced and unforced simulations,
respectively. In this paper, to compare the two pore models, we have
used $m = 1$, and $k T = 1.2$ as in~\cite{luo_slowfast}, unless
otherwise noted. Here, the time step is typically $0.001$.

The number of beads $N$ is odd for polymers initially placed halfway inside the 
pore and even for polymers having initially only the first bead(s) inside the 
pore, see Fig.~\ref{pore_fig}. Before allowing translocation a polymer 
is let to relax for longer than its Rouse relaxation time. We register events 
when a segment $s=s_0$ in the pore is replaced by the segment $s_0-1$ or 
$s_0+1$. The polymer is considered translocated, when it has completely exited
the pore to the {\it trans} side. Exit to either {\it trans} or {\it cis} side 
is regarded as an escape from the pore.

\section{Relating the computational and the physical pore force}
\label{phys_par}

The external driving force, called 
the total pore force, $f_{tot}=nf$, where $n$ is the number of polymer beads
inside the pore and $f$ the $z$-directional driving force excerted on each
bead inside the pore, will be taken as the effective pore force in our model.
The thermal
energy is set to $kT = 1.2$ in reduced units and the length scale is set by the 
polymer bond length $b$. The magnitude of the effective pore force is then 
determined with respect to thermal fluctuations by comparing $f_{tot}b$ 
with $kT$. It is noteworthy, however, that since the pore length is of the 
order of the polymer bond length, there exists an inevitable computational 
artefact due to the discrete polymer model. The external force $f$ is excerted 
on beads
residing in the pore, whose number $n$ does not stay constant during 
translocation. In addition this number is different for different pore lengths, 
so the pore force does not increase strictly linearly with the pore length. These 
effects are, however, mitigated by the fact that $n$ averaged over 
translocation time is constant. 

Correspondence between computational and physical length scale can be 
established by taking the polymer bond length $b$ as the Kuhn length for the 
physical polymer. In SI units the bond length for our FJC model polymer can be 
obtained as $\tilde{b}=2\lambda_p$, where $\lambda_p$ is the persistence 
length, $40\ \text \AA$ for a single-stranded (ss) DNA~\cite{tinland}.
The total pore force in SI units, $\tilde{f}_{tot}$, is then obtained from the 
dimensionless total pore force, $f$, by relating 
$\tilde{f}_{tot} \tilde{b}/k_B \tilde{T} = f_{tot}b/kT$. Here, $k_B$ is 
the Boltzmann constant, and the physical temperature $\tilde{T}$ is taken to be 
$300\ \text{K}$. Thus the effective pore force of $f_{tot} = 1$ corresponds to 
$\tilde{{f}}_{tot} = 1.2\ \text{pN}$ for a ssDNA. To relate to 
experiments, in the $\alpha$-HL pore a typical pore potential of 
$\approx 120\ \text{mV}$ would correspond to 
$\tilde{f}_{tot} \approx 5\ \text{pN}$ when Manning 
condensation leading to drastic charge reduction is taken into 
account~\cite{meller,sauerbudge}. The used computational pore force magnitudes 
are thus in the physically relevant range~\cite{ourepl}.

\section{Results}
\label{res}
\begin{figure}
\centerline{ \includegraphics[angle=0,width=0.5\textwidth]{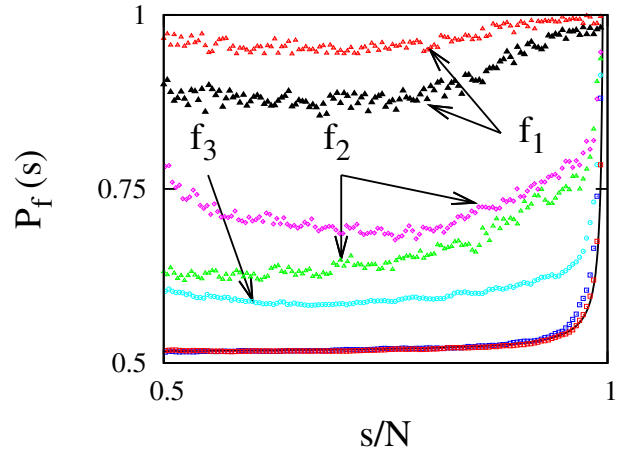} }
\caption[Perturbation force threshold]{
(Color online)
Forward transfer probabilities $P_f$ as functions of the reaction
coordinate normalized with polymer length, $s/N$. The data is given for the 
bead (bd) and cylindrical (cyl) pore. Here $\xi = 0.7$ and
$N$ is $255$ or $256$, depending on the polymer's initial position. 
The pore force $f_{tot}$ has the following values from top to bottom:
$f_1=5.0$ (cyl,bd), $f_2=1.17$ (bd), $f_3=0.5$ (bd).  At the bottom
are $P_f$ for $f_{tot} =0.1$ for the bead (bd, distinct
$\color{blue}\square$ from the solid curve) and the cylindrical (cyl)
pore (red) obtained from simulations together with the black solid
curve calculated from Eq.~(\ref{pf_eq}) for the unforced case. For the 
$f_{tot}$ values $0.1$, $0.5$, and $1.17$ ($f_2$
upper curve) the polymer was initially placed halfway through the pore
$s=(N-1)/2$. For the $f_{tot}$ values $5.0$ (both) and $1.17$ 
($f_2$ lower curve), the
polymer started from the {\it cis} side $s=1$. The shape of the
probability curve depends of the pore model ($f_1$), polymer's initial
position ($f_2$), and changes with the force.
}
\label{transfer_raw}
\end{figure}
\begin{figure*}
\centerline{ \includegraphics[angle=0, width=0.5\textwidth]{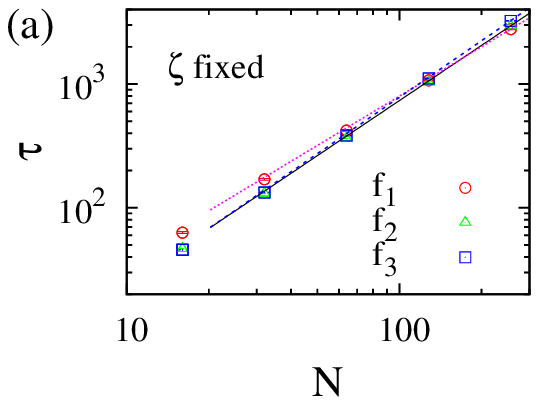}
\includegraphics[angle=0, width=0.5\textwidth]{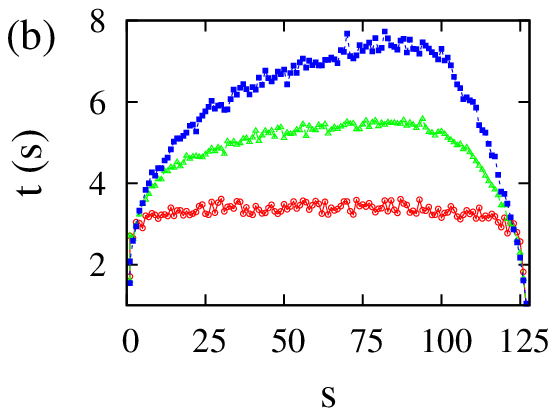} }
\caption{ 
(Color online)
Results from 3D Langevin dynamics with a bead pore model. Keeping 
$\zeta = f_{tot}/ \xi$ fixed, results for three pairs of parameter values $(f_{tot},\xi)$: $(1.17, 0.7)$, $(5, 3)$, and $(10, 6)$, are shown. 
$f_{tot}=f_1$, $f_2$, and $f_3$, respectively.
(a) Scaling of the average translocation time $\tau$ with respect to
the chain length $N$, $\tau \sim N^\beta$. The scaling exponent
$\beta$ has values $1.32 \pm 0.03$ (dotted magenta line), $1.48 \pm
0.03$ (solid black line) and $1.52 \pm 0.03$ (dashed blue line) for
the parameter pairs, respectively.  The exponents are from fits for $N
\in [32,64,128,256]$. 
(b) Average transition times $t(s) = t(s-1 \to s)$ for $N=128$. Plots from 
bottom to top are for $f_{tot}= f_1$, $f_2$, and $f_3$, respectively.
The transition time profile changes significantly when increasing the pore 
force eventhough the proposed contol parameter $\zeta$ is kept fixed.
}
\label{control_parameter}
\end{figure*}
\subsection{Forward transfer probability}
Previously~\cite{ourunf}, we have shown that during the unforced translocation 
the polymer remains close to equilibrium so that the forward transition
probabilities
\begin{equation}
P_f(s) \propto \Big( 1 - \frac{1}{s} + \frac{1}{N-s} \Big)^{1-\gamma}
\label{pf_eq}
\end{equation}
derived from the free energy apply in 3D, where $\gamma = 0.69$. Including the 
pore force in the calculation only introduces a prefactor to the above $P_f(s)$
not changing its form, {\it i.e.} dependence on $s$. For this to hold, the 
translocating polymer has to remain close to equilbrium.
Indeed, except for a small increase due to the force term, the shapes of the 
measured $P_f(s > N/2)$ for the unforced and forced 
translocation with the total pore force $f_{tot} = 0.1$ are 
seen to be similar, cf. Fig.~\ref{transfer_raw}. $P_f$ for the two pore models 
do not differ 
appreciably for this small pore force. Hence we conclude that the transition 
probabilities can be obtained from equilibrium framework for pore force 
magnitudes of the order of $0.1$. 

For the bead pore already for the force $f_{tot}=0.5$, the $P_f(s)$ curve 
differs from 
the equilibrium shape, as shown in Fig.~\ref{transfer_raw}. Hence, the effect 
of the pore force can no more be taken as a small perturbation to equilibrium 
dynamics.
In keeping with this, the transition probabilities $P_f(s > N/2)$ for 
$f_{tot}=1.17$ are seen to depend on the polymer's initial position, here 
either on the {\it cis} side or halfway through the pore. In 
other words, the transition probability from the current state $s$ to the next 
depends on the path through which this state was arrived at. The polymer 
cannot then have relaxed to equilibrium between previous transitions.

Using the cylindrical pore, the probability $P_f$ is seen to be almost
constant and close to one for all $s$ with $f_{tot}=5$, see
Fig.~\ref{transfer_raw}.  This is in accordance with our previous
results that the polymer was seen to almost always translocate with
$f_{tot} \simeq 3$, which means that also in this case $P_f(s>0)$
would be close to one~\cite{ourepl}. However, $P_f(s)$ for the bead
pore deviates significantly from $P_f(s)$ for the cylinder
pore when $f_{tot} = 5$, see Fig.~\ref{transfer_raw}. Thus it is
evident that the pore force magnitudes in the two model pores do not
have an exact correspondence for large pore force magnitudes.

The pore force at which translocation is seen to be a strongly 
out-of-equilibrium process is surprisingly low for both pore models. It may be 
compared with the random force in the one-particle Langevin equation satisfying 
fluctuation-dissipation theorem in 3D, for which 
$\langle f(t)f(t') \rangle = 6 kT \xi \delta(t-t')$. With the parameter 
values $\xi = 0.7$ and $kT = 1.2$ used in our simulations, this yields 
$f \simeq 2.2$.

Already at fairly moderate pore force values, a local force balance
governs the translocation dynamics. This was first proposed by
Sakaue~\cite{sakaue}, and demonstrated in simulations by
us~\cite{ourepl,ourpre}. Considering a concept of `mobile beads' near
the pore, we obtained and described qualitatively the scaling relation 
$\tau \sim N^{1+\nu-\sigma}/f_{tot}$, where the parameter $\sigma$ taking into 
account the varying number of mobile beads diminishes toward zero with 
increasing pore force~\cite{ourepl,ourpre}. Sakaue has recently given a fairly 
quantitative out-of-equilibrium formulation for the scaling 
$\tau \sim N^{1+\nu} /f_{tot}$~\cite{sakaue2010}. However, neither description 
takes into account the crowding of monomers on the {\it trans} side shown to be 
significant for these pore force magnitudes~\cite{ourepl}.

Here, we present results from simulations of the polymer translocation covering 
the relevant range of the pore force. When $f_{tot}b/kT \lessapprox 0.1$, 
$P_f(s)$ attains the equilibrium shape and when 
$f_{tot}b/kT \gtrapprox 5$, $P_f(s)$ is close to one.

\subsection{The proposed control parameter}

Using the bead pore, we first measured the scaling exponent $\beta$ for 
different values of pore force $f_{tot}$ while keeping the proposed control 
parameter $\zeta = f_{tot} / \xi$ fixed. For the parameter pair 
$(f_{tot},\xi)$ values $(1.17,0.7)$, $(5,3)$, and $(10,6)$ were used, see 
Fig.~\ref{control_parameter}~(a). Similar average translocation times $\tau$ 
were obtained for large pore force magnitudes, $f_{tot}=\{5, 10\}$. However, 
$\tau$ differs appreciably for $f_{tot}=1.17$, resulting in a different value 
for the scaling exponent $\beta$.

The measured average waiting (or state transition) times $t(s) = t(s-1
\to s)$ are compared in Fig.~\ref{control_parameter}~(b). The shown three plots 
for $t(s)$ are for the above-mentioned values of $(f_{tot},\xi)$.
In the limit $f_{tot}b/kT \ll 1$, the average state transition 
times $t(s)$ were 
seen to be similar. This is in accord with the transition times in the unforced 
translocation obeying a close-to-Poissonian 
distribution~\cite{ourunf}. When $f_{tot}b/kT \gg 1$ the friction 
of the 
solvent dominates over the stochastic term in the Langevin equation, 
Eq.~(\ref{langevin}). The transition time profiles for the bead pore are 
qualitatively similar with those obtained for the cylindrical 
pore with different parameters~\cite{ourpre}.

From the different scaling of the translocation times obtained for large and 
small pore force magnitudes and the different average state transition times 
we conclude that $\zeta$ which was kept fixed cannot be a universal control 
parameter for forced translocation regardless of the pore model. This is in 
keeping with our previous finding that the forced translocation is a highly 
out-of-equilibriun, non-universal process for the biologically
and experimentally relevant pore force range~\cite{ourepl,ourpre}.

\begin{figure*}
\centerline{
\includegraphics[angle=0,width=0.5\textwidth]{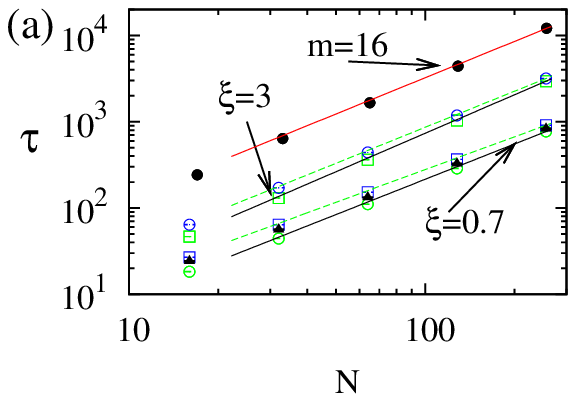}
\includegraphics[angle=0,width=0.5\textwidth]{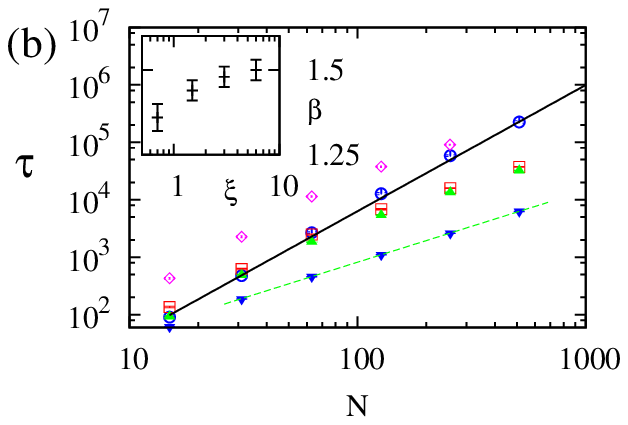}
}
\caption{
(Color online)
Scaling of the average translocation time with respect to the
chain length, $\tau \sim N^\beta$. The results are from 3D Langevin dynamics 
simulations with both cylindrical (cyl) and bead (bd) pore models with 
different pore lengths $l = n b$.
(a) The constant total pore force $f_{tot} = 5$. The exponents are
from fits for $N \in [32,64,128,256]$. For $\xi = 0.7$, the scaling
exponent $\beta= 1.26 \pm 0.02 ({\color{blue}\square})$, $1.27 \pm
0.02 (\blacktriangle)$ and $1.36 \pm 0.03 ({\color{green}\bigcirc})$, for
$l=3$ (cyl), $l=1$ (cyl) and $l=1$ (bd), respectively. For $\xi = 3$,
the scaling exponent $\beta= 1.39 \pm 0.02 ({\color{blue}\bigcirc})$
and $1.48 \pm 0.02 ({\color{green}\square})$, for $l=3$ (cyl) and
$l=1$ (bd), respectively. Increasing the bead mass from $1$ to
$16$ increases the absolute value of $\tau$, and
$\beta=1.40 \pm 0.03 (\bullet)$ for $\xi=0.7$ (cyl). See text for details. 
(b) For $f_{tot} = 0.5$, $\xi = 0.7$ (${\color{blue}
\blacktriangledown}$: bd) the polymers are initially halfway through
the pore. We obtain $\beta = 1.25 \pm 0.04$ (green dashed line) with
$N \in \{31, 63, 127, 255, 511\}$. For $f_{tot} = 0.1$, $\xi = 0.7$
($\color{red} \square$: cyl, ${\color{green} \blacktriangle}$: bd)
polymers are also initially halfway through the pore. The absolute
values of $\tau$ differ slightly for the two the pore models with
shorter polymer chains but not so for $N\simeq 511$ when all chains
escape to the {\it trans} side.  For a very low force $f_{tot} = 0.01
({\color{blue}\bigcirc}$: bd) even the longest ($N=511$) chains may
escape to the {\it cis} side. We obtain $\beta = 2.2 \pm 0.1$ (solid
black line). For reference, results with $f_{tot} = 0.1, \xi = 6$
(${\color{magenta} \diamond}$: bd) are shown.
Inset: The scaling exponent $\beta$ as
a function of the friction constant $\xi$, $f_{tot} = 5$ (bd).
For $\xi = \{0.7, 1.5, 3.0, 6.0\}$, we have $\beta = 1.36 \pm 0.04, 1.44
\pm 0.03, 1.48 \pm 0.03$, and $1.50 \pm 0.03$, respectively.
} 
\label{model_scaling}
\end{figure*}
\subsection{Pore model matters in forced translocation}
\label{pmod}

Originally, the need to compare the cylindrical and bead pore models arose 
from the differences in the scaling of the translocation time $\tau$ with the 
polymer length $N$ in forced translocation obtained using these two pore models.
The translocation times averaged over at least $1000$ and at most $2500$ runs 
are
shown in Fig.~\ref{model_scaling}~(a) as functions of $N$. The total pore force 
is constant, $f_{tot} = 5$, and chosen to be within the experimentally 
relevant range~\cite{ourepl}. For the friction parameter $\xi$ three values, 
$0.7$, $3$, and $6$, are used. For the two separate pore models we find clear 
differences in both the absolute value of $\tau$ and its apparent scaling. 
Regardless of the value for $\xi$, we obtain a smaller scaling exponent 
$\beta$ for the cylindrical than for the bead pore. In addition, increasing 
$\xi$ from $0.7$ to $6$ increases $\beta$ for both pore models, see inset in 
Fig.~\ref{model_scaling}~(b). This in part 
explains the different results obtained with the two pores as the effective 
friction experienced by a chain is different inside them. This difference is 
naturally enhanced at higher velocities.

Our previous results were obtained with the polymer bead mass $m=16$ due to 
the requirement of the stochastic rotation dynamics that the  polymer beads 
should be heavier than solvent beads~\cite{ourepl,ourpre}. Therefore, we 
checked how the bead 
mass affects the characteristics of the translocation time. Increasing the bead 
mass from $1$ to $16$ is seen not only to increase the average 
translocation time $\tau$ but also to change its apparent scaling with polymer 
length $\tau \sim N^\beta$, see Fig.~\ref{model_scaling}~(a). To be certain, we 
checked this using two independent Langevin algorithms.

In accordance with our previous findings for cylindrical pore, $\beta$ was seen 
to increase with pore force also for the bead pore. For 
$f_{tot} = \{0.5, 1.17, 5, 10 \}$, the scaling exponent $\beta =
1.25 \pm 0.04$ (Fig.~\ref{model_scaling}~(b)), $\beta = 1.32 \pm 0.03$
(Fig.~\ref{control_parameter}~(a)), $\beta = 1.36 \pm 0.03$
(Fig.~\ref{model_scaling}~(a)), and $\beta = 1.37 \pm 0.03$ (not
shown) were obtained, respectively. Surprisingly, for the cylindrical pore 
$\beta$, which increased substantially with increasing $f_{tot}$ when 
$m=16$~\cite{ourepl,ourpre}, did not show such strong tendency when $m=1$. 
Changing $m$ seems then to change even the qualitative characteristics of 
forced polymer translocation.

The change of $\beta$ in the scaling $\tau \sim N^\beta$ due to a 
change in $\xi$ is hardly surprising. However, $\beta$ changing with $m$ 
is a bit more subtle. It suggests that we are actually looking at the 
forced translocation in a continually changing or transient stage. This is in 
accord with our previous finding that for polymers of lengths that are used in 
simulations, $N \le 1000$, the number of moving polymer beads changes 
throughout the translocation and thus continually alters the force balance 
condition resulting from the forced translocation taking place out of 
equilibrium~\cite{ourepl}. Also the crowding of the polymer beads on the 
{\it trans} side, whose relaxation toward equilibrium is slower than the 
translocation rate changes the force balance continually.
Hence, the simulated 
forced translocation processes do not reveal the asymptotic ($N > 10^6$) 
characteristics for experimentally relevant pore force magnitudes. Then 
reporting scaling exponents for such forced translocation seems unwarranted. 
The remaining question of interest then is, why does the forced translocation 
exhibit the scaling albeit with varying exponents?

\subsection{Translocation at low pore force}
\label{lowf}

For simulations at low pore force the chains were initially
placed halfway through the pore unlike at large pore force where the
polymer was placed so that only its end was inside the pore.  At the
very low force of $f_{tot}=0.1$ the pore model was found to affect only slightly
the absolute value of $\tau$, see
Fig.~\ref{model_scaling}~(b).  For the bead pore
model chains of length $N \le 127$ had a finite probability to escape
to the {\it cis} side. Unlike for the bead pore, even some of the
$255$ beads long polymers escaped to the {\it cis} side for the
cylindrical pore. This is probably due to the cylindrical pore
aligning the polymer chain toward the pore axis in the middle thus
effectively reducing the friction between the polymer and the pore.

For the pore force $f_{tot} = 0.01$ we obtain $\beta = 2.2 \pm 0.1$,
see Fig.~\ref{model_scaling}~(b), which is the expected exponent for
unforced translocation, $\beta=2\nu+1=2.2$, since $\nu=0.6$ has been
measured for the swelling exponent in our model~\cite{ourpre}.
Approaching the unforced case makes the translocation dynamics increasingly 
robust to variations in the model, which is a natural consequence
of the small pore force not dominating over entropic forces.
We regard the pore force magnitude $0.1$ already small enough 
for the translocating polymer to remain close to equilibrium, see 
Fig.~\ref{transfer_raw}. However, we did not obtain a clear scaling 
for $f_{tot}=0.1$ for either pore models, but $\beta$ close to the unforced 
translocation value $2\nu+1$ was obtained for chains shorter than $N=127$ while 
a lower value of $\beta$ was obtained for longer chains.
This applies for both $\xi=0.7$ and $\xi=6$, see Fig.~\ref{model_scaling}~(b). 
We expect the lower value of $\beta$ to be $1+\nu$ (see~\cite{kantor}), although 
a wider range of $N$ would be needed to confirm this. This apparent cross-over 
behavior for low pore force is outside the scope of the present paper and 
calls for a separate, more thorough investigation.

For the presumably small pore force $f_{tot}=0.5$ we obtained 
$\beta=1.25 \pm 0.04$. This differs considerably from $\beta = 1.58 \pm 0.03$ 
reported in~\cite{luo_slowfast}. From Fig.~\ref{transfer_raw} it can be seen 
that $P_f$ for $f_{tot}=0.5$ clearly deviates from the close-to-unforced form 
of $P_f$ for $f_{tot}=0.1$. 
Together with the low value of $\beta$ this observation suggests that 
$f_{tot}=0.5$ is large enough to drive the polymer more and more out of 
equilibrium as the translocation proceeds. The different values for $\beta$ 
here and in~\cite{luo_slowfast} also support this view, since the two 
simulations started from different initial positions. When the relaxation time 
of the polymer toward thermal equilibrium is slower than the translocation time,
the process possesses memory, or correlation over time, {\it i.e.} the memory 
function $M(t-t_0) \neq \delta(t-t_0)$. Thus, a polymer starting half-way 
through the pore is in a different state than a polymer that has arrived at 
this position and started from another initial position.

The data presented for different pore force magnitudes in Sections~\ref{pmod} 
and \ref{lowf} show that for $f_{tot} \ge 0.5$ the value of $\beta$ increases with 
increasing pore force~\cite{ourepl,ourpre} also for the bead pore.  
For very low pore force $\beta$ is found independent of the pore force.

\section{Summary}
\label{sum}

In summary, we have simulated forced polymer translocation in 3D by using 
Langevin dynamics. 
We have implemented two pore models in our algorithm, (i) a pore 
surrounded by eight immobile beads, which we call bead pore and (ii) a cylindrical 
pore where  a damped harmonic potential confines the beads inside the pore 
region. The present study compares the effect these two pores have on the 
forced translocation.

We measured the forward transition probabilities $P_f(s)$ to determine
the range of pore force that would be sufficiently large to include
cases where polymer translocation takes place close to and strongly
out of equilibrium. We found that the polymer remains close to
equilibrium when the total pore force $f_{tot}b/kT \lessapprox
0.1$. Here the translocation dynamics was found to be robust to
variations in the translocation model, such as the details of the
pore. $P_f(s)$ was found to deviate significantly from the
close-to-equilibrium form already for a pore force as small as $0.5$.
The polymer was found to be driven far from equilibrium when $f_{tot}b/kT 
\gtrapprox 5$.

For small pore force magnitudes the forced translocation processes are 
identical for the two pore models. However, the translocation characteristics 
were found to be increasingly model dependent when 
the pore force is increased. 
This is a natural consequence of the dynamics of the forced translocation being 
determined by a continually changing force-balance condition when the pore 
force is large enough, {\it i.e.}  ($f_{tot}b/kT \gtrapprox 0.5$). Accordingly, 
it seems 
that universal exponents for the forced translocation cannot be found in the 
biologically relevant pore force regime. In addition, attempts to define a 
control parameter whose magnitude would consistently determine these exponents 
would seem futile, which we showed for one proposed candidate by using the bead 
pore. Qualitatively the forced translocation exhibited similar characteristics 
with both pore models, most notably the increase with the pore force of the 
exponent determining the relation between the average translocation time and 
the polymer length. In our view this 
is another indication of the continually changing 
force-balance condition governing the highly non-equilibrium forced 
translocation process.

\begin{acknowledgments}
One of the authors (V.V.L.) thanks Dr. K. Luo for email exchanges.
This work has been supported by the Academy of Finland (Project No.~127766).
The computational resources of CSC-IT Centre for Science, Finland, are 
acknowledged.
\end{acknowledgments}


\end{document}